\documentclass[preprint,showkeys,aps,prb]{revtex4}
\usepackage{amsmath}
\usepackage[dvips]{graphicx}
\usepackage{setspace}
\begin{document}

\title{
The $\phi^4$ Model, Chaos, Thermodynamics, and the\\
 2018 SNOOK Prizes in Computational Statistical Mechanics
}

\author{
William Graham Hoover and Carol Griswold Hoover   \\
Ruby Valley Research Institute                    \\
Highway Contract 60 Box 601                       \\
Ruby Valley, NV 89833                             \\
}

\date{\today}


\vspace{0.1cm}

\begin{abstract}
The one-dimensional $\phi^4$ Model generalizes a harmonic chain with nearest-neighbor Hooke's-Law
interactions by adding quartic potentials tethering each particle to its lattice site. In their
studies of this model Kenichiro Aoki and Dimitri Kusnezov emphasized its most interesting feature :
because the quartic tethers act to scatter long-wavelength phonons, $\phi^4$ chains exhibit Fourier
heat conduction. In his recent Snook-Prize work Aoki also showed that the model can exhibit chaos on the
three-dimensional energy surface describing a two-body two-spring chain. That surface
can include {\it at least two} distinct chaotic seas. Aoki pointed out that the model typically exhibits
{\it different} kinetic temperatures for the two bodies.  Evidently few-body $\phi^4$ problems merit
more investigation. Accordingly, the 2018 Prizes honoring Ian Snook (1945-2013) will be awarded
to the author(s) of the most interesting work analyzing and discussing few-body $\phi^4$ models
from the standpoints of dynamical systems theory and macroscopic thermodynamics, taking into
account the model's ability to maintain a steady-state kinetic temperature gradient as well as
at least two coexisting chaotic seas in the presence of deterministic chaos.
\end{abstract}

\maketitle

\begin{figure}[h]
\includegraphics[width=2.3in,angle=-90,bb=101 16 509 774]{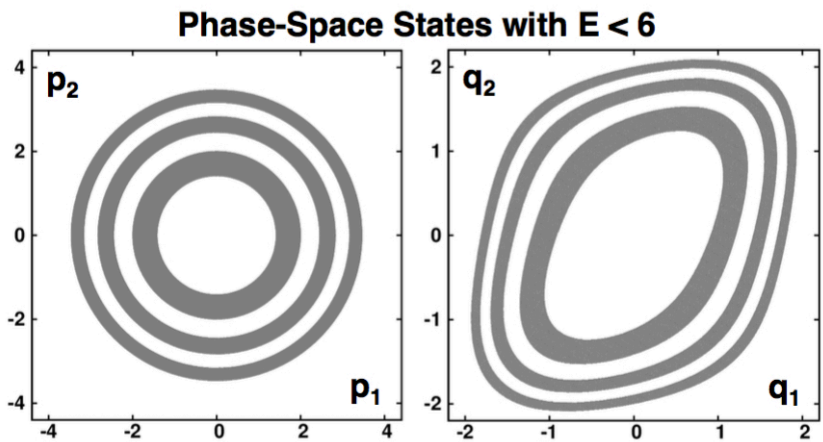}
\caption{
When the two-body $\phi^4$ model has an energy of 6, the momenta are confined to the region $p_1^2 +
p_2^2 < 12$ shown at the left. The displacement coordinates of the particles, $q_1$ and $q_2$,
 are confined to the region shown to the right.  The contours shown here correspond to the energies
1 through 6.
$
E = [ \ p_1^2 + p_2^2 + q_1^2 + (q_1-q_2)^2 \ ]/2  + ( \ q_1^4 + q_2^4 \ )/4 < 6 \ .
$
Most of the three-dimensional microcanonical energy shell between $E=6$ and $E=6+dE$ corresponds to
stable tori.
}
\end{figure}

\section{The Simplest $\phi^4$ Chain and the 2018 SNOOK Prizes}

The 2017 Snook Prize has already shed considerable light on small-system implementations of
Kenichiro Aoki and Dimitri Kusnezov's $\phi^4$ Model\cite{b1}. Besides providing transparent
time-reversible examples of nonequilibrium heat flows the model illustrates several varieties
of broken symmetries in both space and time, as discussed elsewhere in this issue of Computational
Methods in Science and Technology.\cite{b2,b3} {\bf Figure 1} shows equally-spaced contours of
the kinetic and potential energies of the model.

For simplicity, in this work we take initial conditions where the energy is entirely kinetic,
$q_1 = q_2 = 0 \ ; \ p_1^2 + p_2^2 = 12$.  The examples here correspond to the same energy states
studied by Aoki and illustrated in Figures 6 and 7 of his prize-winning contribution for last year's
Snook Prizes\cite{b2,b3}.

In that same competition Timo Hofmann and Jochen Merker discovered two {\it coexisting} chaotic seas
in a fourteen-term polynomial generalization of the H\'enon-Heiles model's cubic Hamiltonian\cite{b4}.
In our follow-up exploration of the two-body $\phi^4$ model we have found two coexisting chaotic seas.
Specimens of both are shown in {\bf Figures 2 and 3}.  Evidently the present
simplest of chaotic Hamiltonians, with only seven polynomial energy contributions, is enough to support
the coexistence of the seas.

\vspace{1.5 cm}
\begin{figure}[h]
\includegraphics[width=3.0in,angle=90,bb= 103 200 514 614]{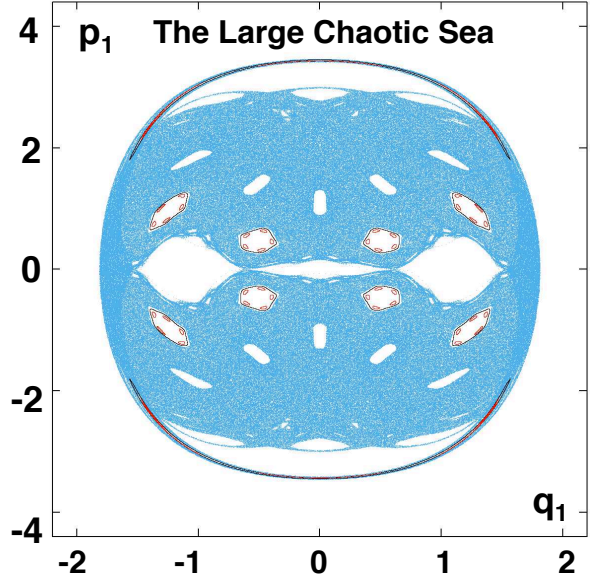}
\caption{A projected section of the ``Large'' sea generated with initial conditions $(q_1,q_2) = (0,0)$
and $(p_1,p_2) = (\sqrt{12},0)$ is shown in blue.  Most of the phase space at this energy corresponds
to tori.  The two examples shown here correspond to initial momenta of $(\sqrt{(11.9,0.1)}$ and
$(\sqrt{(11.8,0.2)}$, with each point on the closed curves plotted when the trajectory passes through
the $q_2= 0$ hyperplane. 
}
\end{figure}

\begin{figure}[h]
\includegraphics[width=3.0in,angle=-90,bb=8 101 593 683]{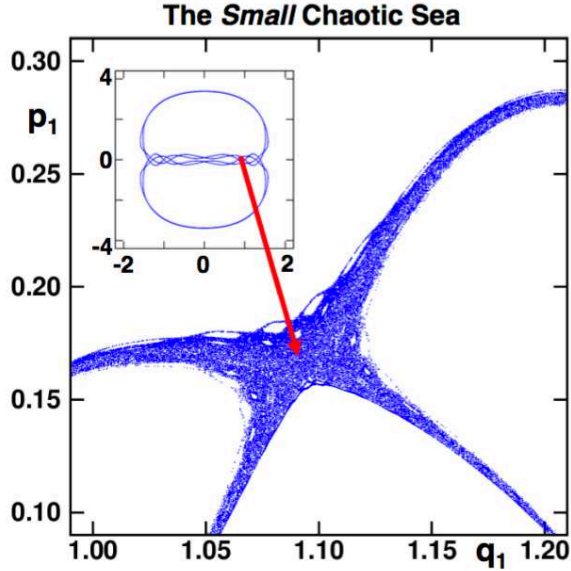}
\caption{
Here the initial condition is  $(q_1,q_2) = (0,0)$ and $(p_1,p_2) = (\sqrt{(11.4,0.6)}$ and the
projection onto the $(q_1,p_1)$ plane is done whenever $q_2=0$.  The full projection is shown in
the upper left inset, where $| \ q_1 \ | < 2$.  An enlargement shows that the apparent crossing
lines in the inset actually correspond to ``fat fractal'' regions with a nonvanishing Lyapunov
exponent, $\lambda_1 = 0.003_0$, where the simulation was extended for $10^{11}$ timesteps in
order to get a reliable value of the exponent.
}
\end{figure}

\section{Chaos in the Two-Mass $\phi^4$ Chains}
Relatively long calculations with $10^{11}$ timesteps showed that both of the problems solved in
{\bf Figures 2 and 3} are chaotic. We used the same reference trajectory + rescaled-satellite
trajectory algorithm discovered independently by groups in Italy
and Japan\cite{b5,b6}.  The small sea in {\bf Figure 3} corresponds to a Lyapunov exponent of 0.003.
The large sea of {\bf Figure 2} is much less stable, with a time-averaged exponent $\lambda_1 = 0.05$.
We wish to emphasize that these two values correspond to exactly the same energy, 6, and only differ
in the initial values of $p_1$ and $p_2$. The Lyapunov-exponent description of the divergence of two
nearby trajectories is defined by the rate equations $\{ \ \dot \delta = \lambda_1 \delta \ \}$,
where the separation $\delta$ is measured in phase space : 
$$
\delta \equiv \sqrt{\delta_{q_1}^2 + \delta_{q_2}^2 + \delta_{p_1}^2 + \delta_{p_2}^2} \ .
$$
The rescaling algorithm brings the satellite trajectory to the same distance, $\delta \rightarrow
0.00001$, after each timestep.  We use fourth-order or fifth-order Runge-Kutta integrators with
$dt = 0.001$ throughout.

{\bf Figure 4} shows the momenta for a time interval $0< {\rm time}<20$ for the large and small
seas.  It is a little
paradoxical that the less stable large-sea trajectory (at the left, with $\lambda_1 = 0.05$)
apparently explores {\it less} of the $(p_1,p_2)$ region than does the more-stable $\lambda_1 =
0.003_0$ small-sea trajectory.\\

\begin{figure}[h]
\includegraphics[width=2.4in,angle=+90,bb=105 25 515 776]{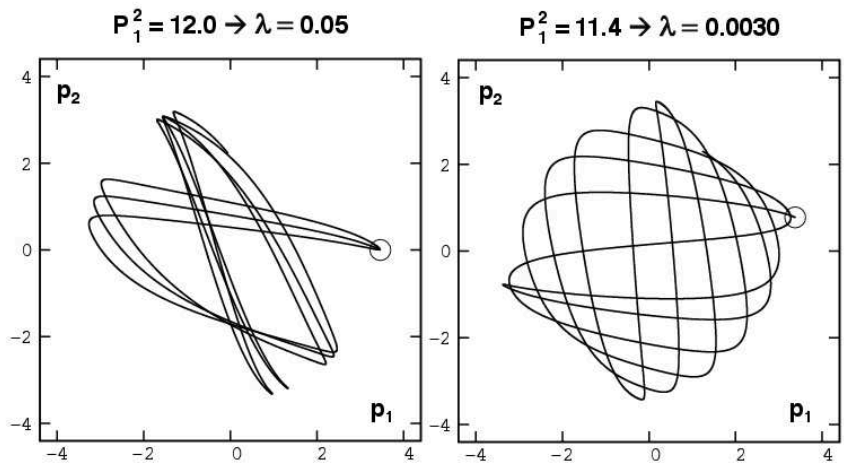}
\caption{
Starting with two circled initial conditions $(p_1^2= 12 \ ; \ p_2^2 = 0)$ and $(p_1^2 = 11.4 \ ;
\ p_2^2 =0.6)$ we show the $(p_1,p_2)$ trajectory projections in momentum space up to a time of 20.
These two trajectories are both chaotic, but with very different Lyapunov exponents.
}
\end{figure}

The three-dimensional energy surface in four-dimensional phase space, $\{ \ q_1,p_1,q_2,p_2 \ \}$ is
difficult to visualize.  Lacking a clever coordinate transformation we can only project or cut.
 Investigation of two-dimensional projections on the six two-dimensional
planes provided by the four state variables shows that much of the surface is composed of tori. For
initial conditions with all or nearly all of the kinetic energy given to Particle 1 at least two
chaotic seas occur.  The sections in {\bf Figure 5} show the chains of islands typical of Hamiltonian
chaos as well as the structures corresponding to simple elliptic doughnuts. It appears that the
chaotic regions correspond to three-dimensional ``fat fractals''\cite{b7}.  The sections provide plenty of
room for further exploration.

\begin{figure}[h]
\includegraphics[width=4.5in,angle=90,bb= 98 185 511 623]{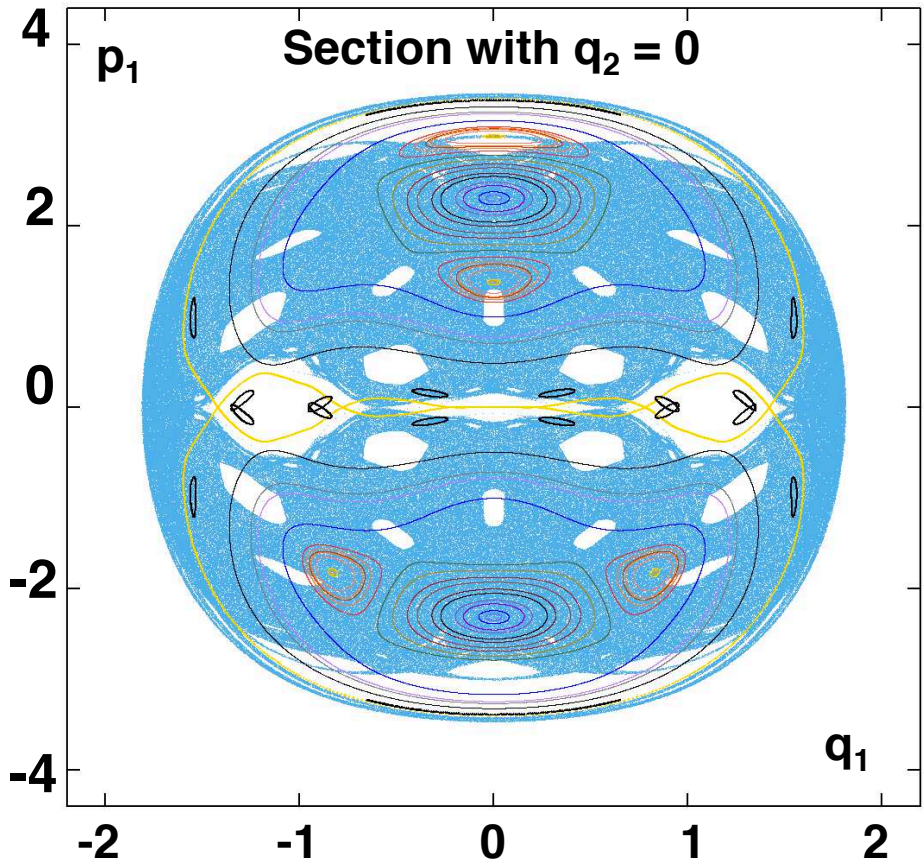}
\caption{
Here we see penetrations of the $(q_1,p_1)$ plane along trajectories of 50,000,000 timesteps
each using 25 equally-spaced initial conditions, $p_1^2 = 0, \ 0.5, \ 1.0, \ 1.5, \ \dots 12$
and $p_1^2 + p_2^2 = 12$. This $q_2 = 0$ projection shows traces of many tori as well as a
black ``chain of islands''.  The last of these initial conditions produces the blue dots, which
form the largest fat-fractal chaotic sea.  Most of the remaining points are closed curves generated by
stable tori.  Note the 18 black curves mostly near $p_1 = 0$ which correspond to a relatively complex
torus which threads through the $q_2=0$ hyperplane eighteen times. The corresponding initial
momenta are $p_1^2 = 11.5 \ ; \ p_2^2 = 0.5$ .
}
\end{figure}

\section{Thermodynamics and the Ideal-Gas Thermometer}

It is interesting to see that the time-averaged kinetic temperatures of the two particles, $T_i =
\langle \ p_i^2 \ \rangle$, are quite different in both the large unstable and the small more stable
chaotic seas.
A permanent temperature difference in a stationary equilibrium system suggests thought-experiments
violating the Second Law of Thermodynamics. Evidently ideal-gas thermometers, though validated by
kinetic theory\cite{b8} cannot be entirely consistent with equilibrium thermodynamics.  This subject
is complicated by the fact that nonequilibrium fractal distributions (typically found for time-reversible
steady states)\cite{b9} correspond to a divergence of the Gibbs entropy $S$, making the usual equilibrium
definition of temperature, $(\partial E/\partial S)_V$, useless.

It is important to see that for any choice of the pair of coordinates $\{ \ q_1,q_2 \ \}$ Gibbs'
statistical mechanics establishes that the maximum-entropy distributions of the two momenta
$\{ \ p_1,p_2 \ \}$ are identical. Thus our finding
$\langle \ T_1 \ \rangle \neq \langle \ T_2 \ \rangle$ shows that the dynamics from Hamilton's motion
equations is not at all ergodic.  For example, in the large chaotic sea the mean values of the kinetic
temperatures of the two particles are roughly $(3.7_4,3.1_7)$.  Whether or not there is a simple and
useful deterministic time-reversible ergodic algorithm for the microcanonical distribution is (we
think) unknown.  Perhaps an analog of magnetic-field rotational forces would be useful in developing
such an algorithm ?

\section{The Snook Prize Problem for 2018}

The several previous $\phi^4$ studies, carried out with a variety of system sizes and thermostatted
boundary conditions\cite{b10}, have established that the $\phi^4$ model can be usefully described by Fourier's
Law. These works also demonstrate that nonequilibrium phase-space distributions are fractal
attractors, with dimensionalities which can lie far below the dimensionality of Gibbs' equilibrium
distributions\cite{b9}. A systematic study could be made to show how the distribution of temperatures
in a conducting chain approaches the Law as the number of degrees of freedom is increased beyond two.
The two-body problem itself suggests a study of the phase-space boundaries separating the
regions of chaos from regular tori and an analysis of the disappearance of the tori with
increasing energy. The possibility of developing a time-reversible ergodic algorithm at constant
energy has to be considered.  A study of clever ideas for the model would be welcome.  The Snook Prize
Problem is a detailed investigation of the two-body $\phi^4$ problem from the standpoints of Hamitonian
chaos and Kolmogorov-Arnold-Moser tori and from the goal of an isoenergetic algorithm for the
microcanonical Gibbs ensemble. It is particularly desirable that Prize entries be self-contained
and pedagogical, stressing numerical findings in sufficient detail that their results can be
corroborated. 

\section{Acknowledgments}

We are grateful to the Institute of Bioorganic Chemistry of the Polish Academy of Sciences and to 
the Poznan Supercomputing and Networking Center for their joint support of these prizes
honoring our late Australian colleague Ian Snook (1945-2013).

\pagebreak

\end{document}